\documentclass{aip-cp}

\usepackage[numbers,elide]{natbib}
\usepackage{graphicx}

\usepackage{listings}
\usepackage[usenames]{color}
\definecolor{codecolor}{gray}{.9}
\definecolor{rlcolor}{cmyk}{0,1,0,0}
\begin{document}

\title{High Density with Elliptic Flows}

\author[aff1]{W. Trautmann\noteref{note1}}
\eaddress{e-mail: w.trautmann@gsi.de}
\affil[aff1]{GSI Helmholtzzentrum f\"{u}r Schwerionenforschung GmbH, Planckstr. 1, 64291 Darmstadt, Germany} 
\authornote[note1]{To appear in the AIP Conference Proceedings of the Xiamen-CUSTIPEN Workshop on the EOS of Dense Neutron-Rich 
Matter in the Era of Gravitational Wave Astronomy (January 3 - 7, 2019, Xiamen, China).}

\maketitle


\begin{abstract}

The elliptic flow of emitted particles and fragments observed in heavy-ion reactions at high energy 
has become an important observable reflecting the pressure generated in the dense collision zone. 
More recently, the strength of the nuclear symmetry energy has been investigated by measuring the ratios or differences 
of the elliptic flows exhibited by neutrons and charged particles in $^{197}$Au+$^{197}$Au collisions 
at 400 MeV/nucleon incident energy at the GSI laboratory. 
A moderately soft to linear dependence on density was deduced for a range of densities shown to reach beyond twice 
the saturation value in these experiments. The known 
sources of uncertainties and possible model dependencies were thoroughly studied with transport models of the UrQMD and T\"{u}bingen 
QMD type. 

A new source of information on the nuclear equation of state at high density has opened up with the observation of 
the first LIGO and Virgo GW170817 gravitational wave signal from a neutron star merger.
The quantitative comparison of terrestrial and celestial results on the basis of measured or inferred neutron star radii
or core pressures, including those obtained from X-ray observations, reveals a rather satisfactory agreement. Depending on 
the precision that can be achieved with future measurements and observations, it will thus become possible to assess the 
validity of the applied models and methods. The perspectives for improved experiments at FAIR using the NeuLAND and KRAB 
detection systems are outlined.

\end{abstract}

\section{INTRODUCTION}
\label{sec:intro}

Calculations indicate that the radii of neutron stars are closely connected to their core pressures at densities up to roughly 
twice the nuclear saturation density~\cite{lattprak2001,lattprak2007}. 
Even a tight correlation between the pressure at saturation density $\rho_0$ and the radius of a neutron star with 1.4 
solar masses has been shown to exist~\cite{newtonli2009,lattprak2016}. 
The pressure range 1-2 $\rho_0$ thus appears as a good interval for comparing terrestrial and 
celestial results for the properties of dense matter~\cite{tsang2018}. 
In the laboratory, heavy-ion reactions are required to produce dense nuclear matter. 
The time interval of the compression affecting mainly the central part of the collision system is short, in the range of 
several to several tens of fm/$c$~\cite{li_npa02,xu13}. Highly valuable and fairly precise results of the pressure vs density 
relation have, nevertheless, been obtained. 
For symmetric matter, pressures for densities of more than four times saturation have been presented, 
deduced from the study of collective phenomena in heavy-ion collisions at incident energies 
up to 10 GeV/nucleon~\cite{dani02}. 
Information on the equation of state of asymmetric matter has more recently been obtained on the basis of differential flows 
measured for $^{197}$Au+$^{197}$Au collisions at 400 MeV/nucleon, a reaction passing through densities up to twice saturation 
according to the calculations~\cite{li_npa02,xu13}. 
The results obtained in two such experiments, their robustness with respect to their interpretation within transport theory, 
and their relation to astrophysical observations are the topic of this talk. 

A few years ago, Li and Han have documented that the many results obtained for the nuclear symmetry energy from terrestrial 
nuclear experiments and astrophysical observations are amazingly compatible, even though individual results scatter within 
considerable margins and are partly affected with large errors~\cite{lihan2013}. The weighted averages calculated by the authors 
for the symmetry energy $E_{\rm sym}$ at saturation density $\rho_0$ and for the slope parameter $L$ describing its density 
dependence are $E_{\rm sym}(\rho_0) = 31.6$~MeV and $L = 58.9$~MeV, respectively. The parameter $L$, defined as

\begin{equation}
L = 3\rho_0 \frac{\partial E_{\rm sym}}{\partial \rho}|_{\rho=\rho_0}, 
\label{eq:L}
\end{equation}

\noindent is the coefficient of the linear term of the Taylor expansion of the symmetry energy with respect to density

\begin{equation}
E_{\rm sym}(\rho)=E_{\rm sym}(\rho_0)+\frac{L}{3}\left(\frac{\rho-\rho_0}{\rho_0}\right)+
\frac{K_{\rm sym}}{18}\left(\frac{\rho-\rho_0}{\rho_0}\right)^2+ ...
\label{eq:Taylor}
\end{equation}

\noindent where $K_{\rm sym}$ is the still fairly unknown coefficient of the quadratic or curvature 
term~\cite{lipr08,liewenchen2019}. The authors quote also error margins representative for the variation of the individual 
results as $\Delta E_{\rm sym}(\rho_0) = 2.7$~MeV and $\Delta L = 16$~MeV and conclude that $L$ has a value about twice as large
as $E_{\rm sym}(\rho_0)$. A very similar conclusion can be drawn from the compilation of Lattimer and Steiner~\cite{lattimer2014} 
adapted from the earlier work of Lattimer and Lim~\cite{lattlim13}. The recent compilation of Oertel {\it et al.} based on 53 
laboratory results and astrophysical observations arrived at nearly the same values $E_{\rm sym}(\rho_0) = 31.7 \pm 3.2$~MeV 
and $L = 58.7 \pm 28.1$~MeV with somewhat different error estimates~\cite{oertel2017}. 

The nuclear symmetry energy governs many aspects of nuclear structure and reactions and determines very basic 
properties and collective modes of neutron stars~\cite{lipr08,epja2014}. 
This implies that many and rather different sources of information exist, in the laboratory and in the cosmos, that provide 
constraints for the equation of state (EoS) of asymmetric nuclear matter. The mentioned comparisons of obtained results
are made possible by the fact that 
the forces identified as best describing a particular observation can be used in many-body calculations to determine
the corresponding $E_{\rm sym}(\rho_0)$ and $L$,  
the two quantities characterizing the asymmetric-matter EoS at and near saturation density. 
It does not require that the measurement or observation has actually tested the EoS at this density. The result may 
represent an extrapolation.  

\begin{figure}[ht]           
  \centerline{\includegraphics[width=290pt]{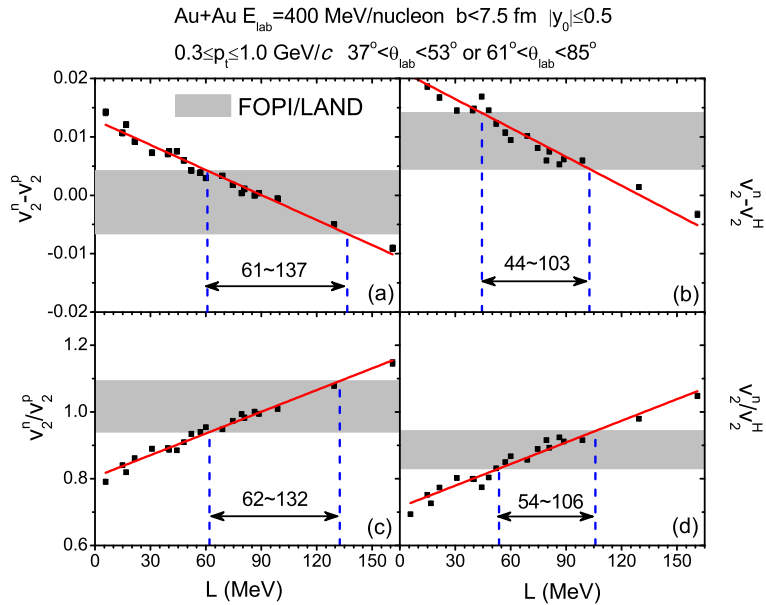}}
   \caption{(Color online) Elliptic flow differences $v_2^{n}-v_2^{p}$ of neutrons with respect to protons (a) and 
$v_2^{n}-v_2^{H}$ of neutrons with respect to hydrogen isotopes (b) and the elliptic flow ratios $v_2^{n}/v_2^{p}$ (c) 
and $v_2^{n}/v_2^{H}$ (d) produced in moderately central $^{197}$Au+$^{197}$Au collisions at 
$E_{\rm lab}=400$~MeV/nucleon as a function of the slope parameter $L$. In each plot, 
the gray shaded region indicates the $p_t/A$-integrated experimental data for the indicated range of laboratory 
angles~\protect\cite{russotto11}, 
full squares denote UrQMD calculations with the selected set of Skyrme forces, while the lines 
represent linear fits to the calculations
(reprinted with permission from Ref.~\protect\cite{wang14}; 
Copyright (2014) by the American Physical Society).}
\label{fig:wang}
\end{figure}

The predictions of microscopic models for the nuclear symmetry energy, obtained with realistic or phenomenological 
forces, appear to coincide at densities close to $\rho = 0.1$~fm$^{-3}$, i.e. at approximately two thirds of the 
saturation density~\cite{fuchs06}. It reflects the fact that the average density of atomic nuclei is below saturation 
and that the presence of the nuclear surface influences the properties that are chosen as constraints. The awareness 
that each observable carrying information on the nuclear EoS is connected to its proper range of density has to 
complement the interpretation of existing results. 

Sensitivities to higher densities can be expected from observables related to the early phases of heavy-ion 
collisions at sufficiently high energies. Calculations predict that the resulting pressure in the central collision zone
produces a collective outward motion, or squeeze-out, of the compressed material whose strength is influenced by the 
symmetry energy in asymmetric systems~\cite{dani02}. A measurement differentiating between the collective flows of 
neutrons and protons can thus be expected to provide information on the high-density symmetry energy~\cite{li02}.
The ratio of the elliptic flow strengths observed for neutrons and light charged particles is one of the observables 
proposed for the study of the EoS of asymmetric matter~\cite{russotto11}. 
It is approximately linearly correlated with the slope parameter $L$ (Eq.~\ref{eq:L}) as shown by Wang {\it et al.}~\cite{wang14}. 
In their work, the differential-flow data available from the FOPI-LAND experiment~\cite{russotto11} were compared with 
calculations performed with the updated version of the
Ultrarelativistic-Quantum-Molecular-Dynamics (UrQMD) model
using 21 selected Skyrme forces. 
The ratios as well as the differences of the elliptic flow parameters were found to exhibit a linear correlation as a 
function of $L$ (Fig.~\ref{fig:wang}). The intervals compatible with the experimental 1-$\sigma$ 
error bands do not depend on whether the ratios or differences are employed. The results are 
slightly different, however, when all hydrogen isotopes (indicated by H) instead of only protons (p) are selected. 
As shown below, these two observables differ somewhat in the density intervals they are sensitive to. 
The intervals of $L$ found compatible with the experiment are large because of the limited statistics collected 
in the FOPI-LAND experiment.

A higher precision was achieved in the ASY-EOS experiment for the same $^{197}$Au+$^{197}$Au reaction~\cite{russotto16}. Also the 
robustness of the differential elliptic flows with respect to their interpretation within transport theory has been investigated 
and established~\cite{cozma13,cozma18}. With the sensitivity to density reaching beyond twice saturation, a comparison with 
results obtained from astrophysical observations becomes feasible. Besides the pressures, also the radii of 1.4 solar mass 
neutron stars are useful quantities for this purpose. For pressures above saturation, the contribution of the incompressibility 
of symmetric matter has to be considered. For the conversion of pressures to radii and vice versa, the correlation established 
in Ref.~\cite{lattprak2016} will be used.

\section{THE FOPI-LAND AND ASY-EOS EXPERIMENTS}
\label{sec:exp}

The FOPI-LAND experiment conducted in 1991 at the GSI laboratory represented the first measurement of the elliptic flow of 
neutrons~\cite{leif93,lamb94}. The  Large-Area-Neutron-Detector (LAND), a new instrument at that time, was one of the two 
main detction systems used in the experiment. LAND is a $2 \cdot 2 \cdot 1$~m$^3$ calorimeter, consisting of in total 200 
slabs of interleaved iron and plastic strips viewed by photomultiplier tubes at both ends, and is used as a time-of-flight 
spectrometer~\cite{LAND}. For the experiment, it had been divided into two 
parts of 0.5-m depth each, positioned side by side with the aim of covering a large range in rapidity 
centered at midrapidity for nucleon transverse momenta up to about 1 GeV/$c$. Veto walls in front of the two sections
permitted recording light charged particles with the same large acceptance. 
During its Phase 1, the 4-$\pi$ detector FOPI consisted of a forward plastic array built 
from more than 700 individual plastic-scintillator strips and covering the forward hemisphere
at laboratory angles from 1$^{\circ}$ to 30$^{\circ}$~\cite{FOPI}. 
It was used to determine the modulus and orientation of the impact parameter from the multiplicity and 
azimuthal distribution of the detected charged particles~\cite{reis97}. 
These data are still available and were reanalyzed in order to determine optimum
conditions for a dedicated new experiment, but also with the
aim to produce constraints for the symmetry energy by comparing with predictions of state-of-the-art transport models. 
The results obtained with the 400-MeV/nucleon data set and the UrQMD model~\cite{qli05} were reported in Ref.~\cite{russotto11}.
A moderately soft to linear symmetry term, characterized by a coefficient 
$\gamma = 0.9 \pm 0.4$ for the power-law parametrization of the density dependence of the potential part of the symmetry 
energy was obtained~\cite{russotto11,cozma11,traut12,russotto_epja14}.
 
\begin{figure}[ht]            
  \centerline{\includegraphics[width=265pt]{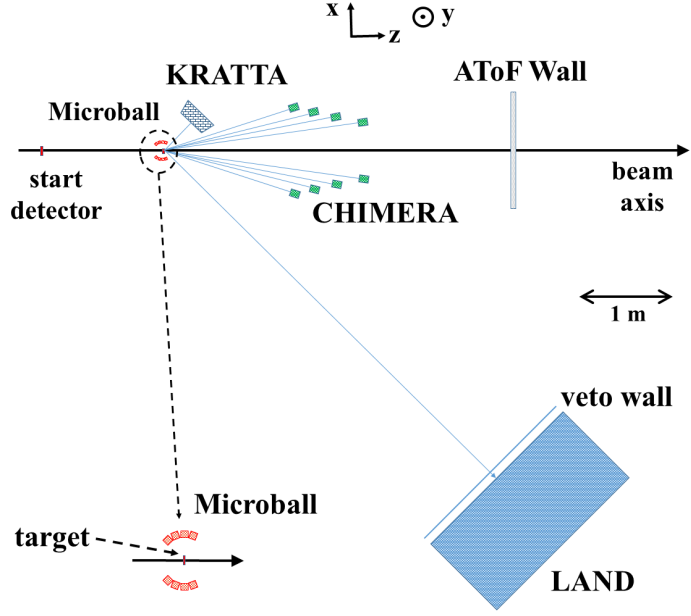}}
   \caption{(Color online) Schematic view of the experimental setup of the ASY-EOS experiment S394 at GSI. 
The chosen coordinate system is indicated,
the $y$ direction points upwards in the laboratory. The target area with the
Microball is not to scale in the main drawing but shown with a scale factor of approximately 5:1 in the lower left corner
(from Ref.~\protect\cite{russotto16}; Copyright (2016) by the American Physical Society).
}
\label{fig:setup}       
\end{figure}

Motivated by this finding, an attempt was made to improve the 
accuracy with a new experiment that was conducted at the GSI laboratory in 2011 (ASY-EOS experiment S394). 
The experimental setup followed the scheme developed for FOPI-LAND by 
using LAND as the main instrument for neutron and charged particle 
detection~(Fig.~\ref{fig:setup}). This time, LAND was used as a single unit of 1-m depth.  
Opposite of LAND, covering a comparable range of polar angles, the Krak\'{o}w Triple Telescope 
Array (KRATTA~\cite{Luk11}) was installed to permit flow measurements of identified 
charged particles under the same experimental conditions. 

For the event characterization and for measuring the orientation of the reaction plane, three
detection systems had been installed. The ALADIN Time-of-Flight (AToF) Wall~\cite{schuettauf96} was used 
to detect charged particles and fragments in forward direction at polar angles up to
$\theta_{\rm lab} \le 7^{\circ}$. Its capability of identifying large fragments and of characterizing
events with a measurement of $Z_{\rm bound}$~\cite{schuettauf96} permitted the sorting of events 
according to impact parameter. Four double rings of the CHIMERA multidetector~\cite{Pag04,DeFilippo14} 
carrying together 352 CsI(Tl) scintillators in forward direction and four rings with 50 thin CsI(Tl) 
elements of the Washington University Microball array~\cite{muball} surrounding the target
provided sufficient coverage and granularity for determining the orientation 
of the reaction plane from the measured azimuthal particle distributions. A detailed description of the 
experiment is available in Ref.~\cite{russotto16}.

\section{ASY-EOS EXPERIMENTAL RESULTS}
\label{sec:urqmd}

\begin{figure}[ht]           
  \centerline{\includegraphics[width=200pt]{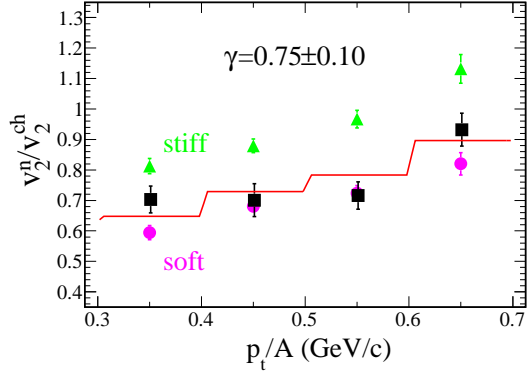}}
   \caption{(Color online) Elliptic flow ratio of neutrons over all charged particles for central ($b<$ 7.5 fm) 
collisions of $^{197}$Au+$^{197}$Au at 400 MeV/nucleon
as a function of the transverse momentum per nucleon $p_{t}/A$. 
The black squares represent the experimental data, the green triangles and purple circles represent the UrQMD predictions
for stiff ($\gamma =1.5$) and soft ($\gamma =0.5$) power-law exponents of the potential term, respectively. 
The solid line is the result of a linear interpolation between the predictions, weighted according to the experimental 
errors of the included four bins in $p_{t}/A$, and leading to the 
indicated $\gamma =0.75 \pm 0.10$
(from Ref.~\protect\cite{russotto16}; Copyright (2016) by the American Physical Society).
}
\label{fig:russotto_fig14}       
\end{figure}

As in the FOPI-LAND experiment, the reaction $^{197}$Au+$^{197}$Au at 400 MeV/nucleon was studied. Nuclear stopping reaches a 
maximum at this incident energy~\cite{andronic06} and the squeeze-out of light charged particles assumes its largest 
values~\cite{lefevre18}. 
The elliptic flows of neutrons and light charged particles were determined from the azimuthal distributions of these 
particles with respect to the reaction plane reconstructed from the distributions of particles and fragments recorded 
with the three arrays AToF, CHIMERA, and Microball. 
Methods developed and described in Refs.~\cite{andronic06,Ollxx} for correcting the finite dispersion of the reaction 
plane were applied. The coefficients $v_1$ and $v_2$ representing the strengths of directed and elliptic flows, 
respectively, were deduced from fits with the Fourier expansion 

\begin{equation}
f(\Delta\phi) \propto 1 + 2 v_1 {\rm cos}(\Delta\phi) + 2 v_2 {\rm cos}(2 \Delta\phi).
\label{eq:fourier}
\end{equation}

\noindent Here $\Delta\phi$ represents the azimuthal angle of the momentum vector of an 
emitted particle with respect to the angle representing the azimuthal orientation of the reaction plane. 
Constraints for the symmetry energy were determined by comparing the ratios 
of the elliptic flows of neutrons and charged particles (ch), $v_2^{n}/v_2^{ch}$, with the corresponding UrQMD 
predictions for soft and stiff assumptions.

For the analysis, the UrQMD model was employed in the version
adapted to the study of intermediate energy heavy-ion collisions~\cite{qli11}.
The chosen isoscalar EoS is soft and
different options for the dependence on isospin asymmetry were implemented. Two of them are used here, 
expressed as a power-law dependence of the potential part of the symmetry energy on the
nuclear density $\rho$ according to

\begin{equation}
E_{\rm sym} = E_{\rm sym}^{\rm pot} + E_{\rm sym}^{\rm kin} 
= 22~{\rm MeV} (\rho /\rho_0)^{\gamma} + 12~{\rm MeV} (\rho /\rho_0)^{2/3} 
\label{eq:pot_term}
\end{equation}

\noindent with $\gamma =0.5$ and $\gamma =1.5$ corresponding to a soft and a stiff density 
dependence, respectively.  

The predictions obtained with these assumptions for the measured flow ratio are shown in Fig.~\ref{fig:russotto_fig14} together 
with the experimental result. The histogram represents the linear interpolation between the predictions giving a best 
fit of the flow ratios presented here as a function of the transverse momentum per nucleon $p_t/A$. The corresponding power-law 
coefficient is $\gamma = 0.75 \pm 0.10$ with a purely statistical error $\Delta \gamma = 0.10$.   
In comparison with FOPI-LAND, this represents an improvement of the statistical accuracy by a factor of four. 

\begin{figure}[ht]           
  \centerline{\includegraphics[width=250pt]{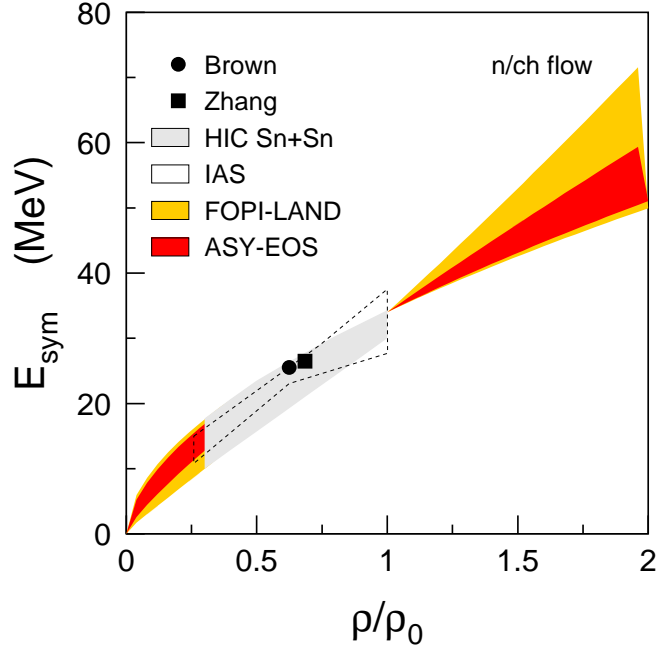}}
   \caption{(Color online) Constraints deduced for the density dependence of the symmetry energy from the ASY-EOS 
data~\protect\cite{russotto16} in comparison with the FOPI-LAND result of Ref.~\protect\cite{russotto11} as a function 
of the reduced density $\rho/\rho_0$. 
The low-density results of Refs.~\protect\cite{brown13,zhang13,tsang09,dani14} as reported in Ref.~\protect\cite{horowitz13} 
are given by the symbols, the grey area (HIC), and the dashed contour (IAS). For clarity, the FOPI-LAND and ASY-EOS 
results are not displayed in the interval $0.3 < \rho/\rho_0 < 1.0$
(from Ref.~\protect\cite{russotto16}; Copyright (2016) by the American Physical Society).
}
\label{fig:russotto_fig18}       
\end{figure}

The necessary corrections and the methods used for estimating systematic uncertainties are described in detail 
in Ref.~\cite{russotto16}. In particular, after applying the necessary timing corrections, the charged particle flows were 
consistent with FOPI data sorted with identical acceptance criteria. With all the corrections and errors arising from 
the analysis included, the acceptance-integrated elliptic-flow ratio leads to a power-law 
coefficient $\gamma = 0.72 \pm 0.19$. This is the result displayed in Fig.~\ref{fig:russotto_fig18} as a function of the 
reduced density $\rho/\rho_0$. The new result, represented by the red band in the figure, confirms the former (yellow band) 
and has a considerably smaller uncertainty. 
The slope parameter describing the variation of the symmetry energy with density at saturation
is $L = 72 \pm 13$~MeV. The corresponding curvature term in the Taylor expansion with respect to density is in the 
range $K_{\rm sym} = -70$ to -40~MeV.

The present parametrization is shown to be compatible with the low-density behavior of the symmetry 
energy from Refs.~\cite{brown13,zhang13,tsang09,dani14} that are included in the figure.  Rather precise values for 
the symmetry energy and for the density to which it applies have recently
been presented by Brown~\cite{brown13} and Zhang and Chen~\cite{zhang13}. They are shown in Fig.~\ref{fig:russotto_fig18} 
together with the low-density behavior of the symmetry energy obtained from heavy-ion collisions~\cite{tsang09} and 
from the analysis of isobaric analog states~\cite{dani14} as reported in Ref.~\cite{horowitz13}. In the study of 
Brown, a set of selected Skyrme forces is used whose parameters are fitted to properties of doubly magic nuclei. 
By using particular values for the neutron skin of $^{208}$Pb nuclei within a given range as additional constraints, 
new sets of of these forces with slightly adjusted parameters are obtained. It is found that all predictions coincide 
at a density $\rho = 0.1$~fm$^{-3}$, independent of the choice made for the thickness of the neutron skin, but that 
the slopes at this density depend on this choice. Only a precise knowledge of the neutron skin of $^{208}$Pb will 
permit the extrapolation to the saturation point. It underlines the importance of this nuclear property determined by the 
balance of pressures experienced by neutrons in the neutron-enriched interior of the $^{208}$Pb nucleus and in the low-density 
neutron-rich surface~\cite{rocamaza11,prex12}.

The crossing of the error bands at the fixed value $E_{\rm sym} (\rho_0) = 34$~MeV is a consequence of the chosen parametrization 
(Eq.~\ref{eq:pot_term}). Using values lower than the default $E_{\rm sym}^{\rm pot} (\rho_0) = 22$~MeV, as occasionally 
done in other UrQMD studies~\cite{wang14a}, will lower the result for $L$. Performing the present UrQMD analysis 
with $E_{\rm sym}^{\rm pot} (\rho_0) = 19$~MeV, corresponding to $E_{\rm sym} (\rho_0) = 31$~MeV, yields a power-law 
coefficient $\gamma = 0.68 \pm 0.19$ and a slope parameter $L = 63 \pm 11$~MeV. The observed changes remain both 
within the error margins of these quantities. However, the precise results of Brown~\cite{brown13} and Zhang and 
Chen~\cite{zhang13} are no longer equally met with this alternative parametrization of the symmetry energy. The consistency of 
nuclear-structure results for the low-density behavior of the symmetry energy appears to be increasing. 
The analysis of the dipole polarizability in $^{208}$Pb performed by Zhang and Chen~\cite{zhang2015} and discussed at this 
conference~\cite{liewenchen2019,lynch2019} confirms the trend shown 
in Fig.~\ref{fig:russotto_fig18} towards densities around $\rho_0 /3$. 

Results from X-ray observations of neutron stars were recently combined with global limits and, with a parametrized set 
of model equations of state, used to generate constraints for the density dependence of the nuclear symmetry 
energy~\cite{zhang_li2018}. The authors show that the value of the symmetry 
energy of $47 \pm 10$~MeV represents a significant constraint that can be obtained from coherently analyzing several 
data sets. It overlaps partly with the $56 \pm 5$~MeV interval given by the UrQMD parametrization of the ASY-EOS result
(Fig.~\ref{fig:russotto_fig18}). Altogether, one may conclude that the obtained parametrization bridges well 
the density interval from low densities up to twice the saturation value, leading to a consistent description of 
the density-dependent symmetry energy,
based on nuclear structure and reaction data as well as on X-ray observations of neutron stars.

\section{SENSITIVITY TO DENSITY}
\label{sec:density}

Together with the presentation of the results~\cite{russotto16}, the range of densities probed with the elliptic-flow 
ratio was explored using the T\"{u}bingen version of the QMD model (T\"{u}QMD, Ref.~\cite{cozma13}). The applied method 
consisted of performing transport calculations for the present reaction with two 
parametrizations of the symmetry energy that were chosen to be different for a selected range of density and identical 
elsewhere. The magnitude of the obtained difference between the two 
predictions for the elliptic flow ratio is interpreted as a quantitative measure of the sensitivity to the selected density 
region.

In the momentum-dependent one-body potential introduced by Das {\it et al.}~\cite{das03}, the stiffness of 
the symmetry energy is controlled with a parameter $x$ and choices ranging from soft ($x=+1$) up to rather stiff 
($x=-2$) density dependences are commonly selected in model studies.
For the present case, a mildly stiff ($x=-1$) and a soft ($x=+1$) parametrization were chosen for the density range 
with different symmetry energies while the nearly linear case with $x=0$ was chosen for the common part. 

To quantify the results, a quantity DEFR (Difference of Elliptic-Flow Ratio) was defined as
\begin{eqnarray}
{\rm DEFR}^{(n,Y)}(\rho)=\frac{v_2^n}{v_2^{Y}}(x=-1,\rho)-\frac{v_2^n}{v_2^{Y}}(x=1,\rho).
\label{defrdef}
\end{eqnarray}
It represents the differences of the elliptic flow ratios calculated with the T\"{u}QMD transport model for two different 
density dependences of the symmetry energy. Here $Y$ indicates a charged particle or a group 
of charged-particle species and $x$ the stiffness parameter that is used in the calculations at densities smaller than 
the argument $\rho$. 
The difference of the $x = \pm 1$ potentials is thus effectively only tested
at densities up to the particular $\rho$, the argument of DEFR. This choice leads 
to DEFR$^{(n,Y)} (0) = 0$ and to the full stiff-soft splitting for large values of $\rho$. The slope of DEFR
at intermediate densities is a measure of the impact on elliptic flow observables of that particular 
region of density. 

\begin{figure}[ht]           
  \centerline{\includegraphics[width=190pt]{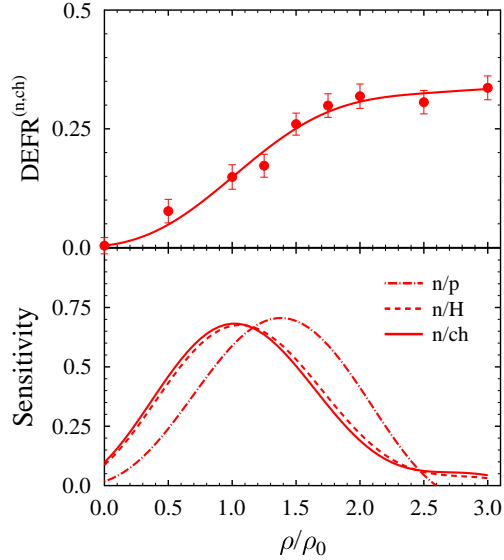}}
   \caption{(Color online) Density dependence of the difference of the elliptic flow ratio (DEFR) of neutrons
over charged particles, defined by Eq.~\protect\ref{defrdef}, for $^{197}$Au+$^{197}$Au collisions 
at 400 MeV/nucleon obtained with the T\"{u}QMD transport model~\protect\cite{cozma13} and 
the FOPI-LAND acceptance filter (top) and the corresponding sensitivity density (bottom panel, solid line) together with
sensitivity densities obtained from elliptic-flow ratios of neutrons over all hydrogen isotopes (dashed) and 
neutrons over protons (dash-dotted; 
from Ref.~\protect\cite{russotto16}; 
Copyright (2016) by the American Physical Society).
}
\label{fig:dens}       
\end{figure}

In the upper panel of Fig.~\ref{fig:dens}, the density dependence of DEFR$^{(n,Y)}$ for the choice $Y$=all charged 
particles is presented. It is seen that DEFR increases monotonically up to density values in the neighborhood 
of 2.5\,$\rho_0$, close to the maximum density probed by nucleons in heavy-ion collisions at 400 MeV/nucleon 
incident energy~\cite{li_npa02,xu13}. The distribution of the sensitivity as a function of density is obtained by forming the 
derivative of DEFR with respect to density.
It is presented in the lower panel of Fig.~\ref{fig:dens} for three choices of $Y$, only protons (n/p), sum of all 
hydrogen isotopes (n/H), and all charged particles (n/ch).
 
The figure shows that the sensitivity achieved with the elliptic-flow ratio of neutrons over charged particles, 
the case of the ASY-EOS experiment, reaches its maximum close to saturation density and extends beyond twice that value. 
It is compatible
with the conclusions reached by Le~F\`{e}vre {\it et al.} in their study of the symmetric matter EoS, based on FOPI 
elliptic-flow data and calculations with the Isospin Quantum Molecular Dynamics (IQMD) transport model~\cite{lefevre16}.
For $^{197}$Au+$^{197}$Au collisions at 400 MeV/nucleon, the broad maximum of the force-weighted density defined by the authors
is spread over the density range $0.8 < \rho /\rho_0 < 1.6$.

The sensitivity of the neutron-vs-proton flow ratio has its maximum in the 1.4 to 1.5 $\rho_0$ region, i.e. at 
significantly higher densities than with light complex particles being included (Fig.~\ref{fig:dens}).
By probing different densities with different ratios, the curvature of the symmetry energy at saturation, 
in addition to the slope, becomes accessible~\cite{russotto16}. As shown below, this was demonstrated by Cozma with very recent 
calculations~\cite{cozma18}. 

\section{PRESSURES}

The value $\gamma = 0.72 \pm 0.19$ obtained in the ASY-EOS experiment for the power-law coefficient of the potential part 
in the UrQMD parametrization of the symmetry energy and the slope parameter $L = 72 \pm 13$~MeV are equivalent to a symmetry 
pressure $p_0 = \rho_0 L/3 = 3.8 \pm 0.7$~MeVfm$^{-3}$. It represents 
the pressure in pure neutron matter at saturation because the pressure in symmetric matter vanishes at this density.
It may be used to estimate the pressure in neutron-star matter at 
saturation density. For an assumed
asymmetry $\delta = (\rho_n - \rho_p)/\rho = 0.9$ in that part of the star~\cite{lipr08,margueron2018} and after adding the 
small contribution of the degenerate electrons~\cite{lipr08}, 
the value $3.4 \pm 0.6$~MeVfm$^{-3}$ is obtained~\cite{russotto16}. 
It is located inside the pressure interval 
obtained with 95\% confidence limit by 
Steiner {\it et al.}~\cite{steiner13} from the observation of eight neutron-stars. It touches into the 50\% confidence 
interval (1.6 to 3.0~MeV/fm$^{-3}$) of the pressure-density relation presented by the LIGO and Virgo 
collaborations~\cite{abbott2018}. 
The 90\% interval extends up to 4.1~MeV/fm$^{-3}$), a value that may be considered an upper limit consistently supported 
by X-ray observations, the analysis of the GW170817 gravitational wave event, and by laboratory measurements using 
heavy-ion reactions.

\begin{figure}[ht]           
  \centerline{\includegraphics[width=320pt]{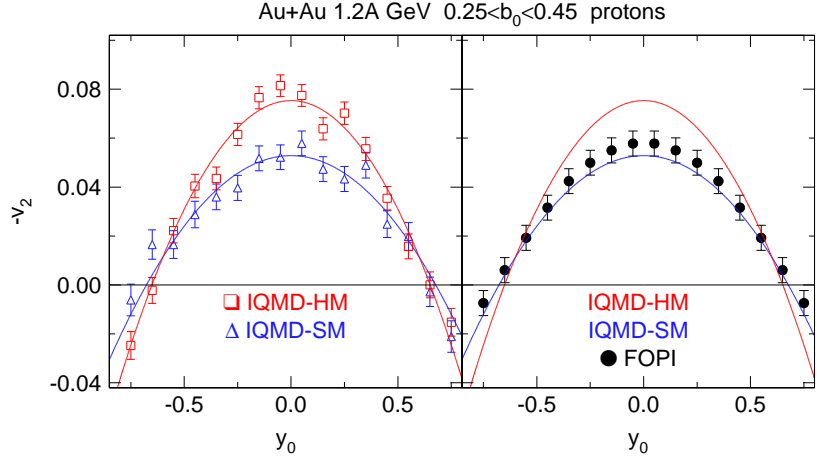}}
   \caption{(Color online) Elliptic flow parameter -v2 (note the change of sign) of protons as a function of 
the normalized rapidity $y_0$
for $^{197}$Au + $^{197}$Au collisions at 1.2 GeV/nucleon and the indicated near-central interval of
normalized impact parameters $b_0$. Left panel: IQMD calculations (symbols) for a hard (HM) and a soft (SM)
EoS, both with momentum dependent forces, and the fit results (lines) assuming a quadratic dependence on $y_o$.
Right panel: The obtained fit results (lines) in comparison with the experimental data (symbols) measured
with the FOPI detector at the GSI laboratory
(reprinted from Ref.~\protect\cite{lefevre16}, Copyright (2016), with permission from Elsevier).
}
\label{fig:lefevre}
\end{figure}

The sensitivity of the ASY-EOS data was shown to extend to about twice the saturation density. To estimate the neutron-star 
pressure at this density requires knowledge of the pressure expected for symmetric matter at the same density. 
The existing FOPI data for directed and elliptic flows of light charged particles 
in $^{197}$Au+$^{197}$Au collisions at incident energies up to 1.5 GeV/nucleon~\cite{reisdorf12} have recently been 
reinterpreted with the IQMD and UrQMD transport models~\cite{lefevre16,wang2018}. Both analyses favor soft solutions with 
momentum dependent interactions. Le~F\`{e}vre {\it et~al.} obtain an excellent description of the rapidity dependence of 
the proton elliptic flow with the IQMD model (Fig.~\ref{fig:lefevre}), but also for deuterons, tritons and $^{3}$He particles, 
and deduce an incompressibility parameter $K_0 = 190 \pm 30$~MeV from their IQMD study~\cite{lefevre16}. Wang {\it et~al.} 
use three selected Skyrme forces to cover a large parameter interval of $K_0$ with the UrQMD model and analyze the functional 
form of the rapidity dependence of the proton 
and deuteron elliptic flows at energies up to 1 GeV/nucleon~\cite{wang2018}. 
They report a fairly similar value $K_0 = 220 \pm 40$~MeV as best 
describing the data. From the averaged result $K_0 = 200 \pm 30$~MeV a pressure $p (2\rho_0) = 14.2 \pm 2.1$~MeV/fm$^{3}$ is 
obtained for symmetric matter with density 2~$\rho_0$. 

Together with the contribution of the symmetry energy at this density, 
based on the UrQMD power-law parametrization, and with an electron pressure of about 1~MeV/fm$^{3}$~\cite{lipr08}, a 
value $p (2\rho_0) = 23 \pm 3$~MeV/fm$^{3}$ is obtained for neutron star matter. 
It is equal to $3.7 \pm 0.5 \cdot 10^{34}$~dyn/cm$^2$, a value falling into the
middle of the 50\% confidence margin of the LIGO and Virgo result~\cite{abbott2018}. The uncertainty is rather small. 
The contributions of the incompressibility of symmetric matter and of the density dependence of the symmetry energy are 
about equal and were added quadratically. Following 
Margueron {\it et al.} \cite{margueron2018}, a 
proton fraction $x_p = 0.1$ is assumed for the density $2\rho_0$. Lowering the proton fraction 
to $x_p = 0.06$~\cite{lattprak2001} raises the pressure by 1.7~MeV/fm$^{3}$ or $0.3 \cdot 10^{34}$~dyn/cm$^2$.
Altogether, the comparisons of neutron star pressures in the 1-2~$\rho_0$ interval can be considered as extremely promising. 
Even more meaningful comparisons can be expected with new data resulting from the observation of new merger events and with 
further progress in stabilizing the transport description as a result of the presently 
ongoing code comparison project~\cite{junxu16,zhang2018}.

\section{TOWARDS MODEL INDEPENDENCE WITH MDI2}

The potential model dependence of the obtained results has been a concern in the analysis and interpretation of 
the measured flow data from the very beginning of these studies. When reporting the FOPI-LAND results, the robustness of the 
elliptic-flow ratio of neutrons over hydrogen isotopes was demonstrated by using two
different parametrizations for the momentum dependence of the in-medium nucleon-nucleon elastic scattering cross 
sections~\cite{russotto11}. They either overestimated or underestimated by typically 15\% the $Z=1$ elliptic flows measured by 
the FOPI Collaboration~\cite{andronic2005}. Their effect on the absolut flow values is thus significant but was shown to 
largely cancel in the flow ratios.

\begin{figure}[ht]           
  \centerline{\includegraphics[width=250pt]{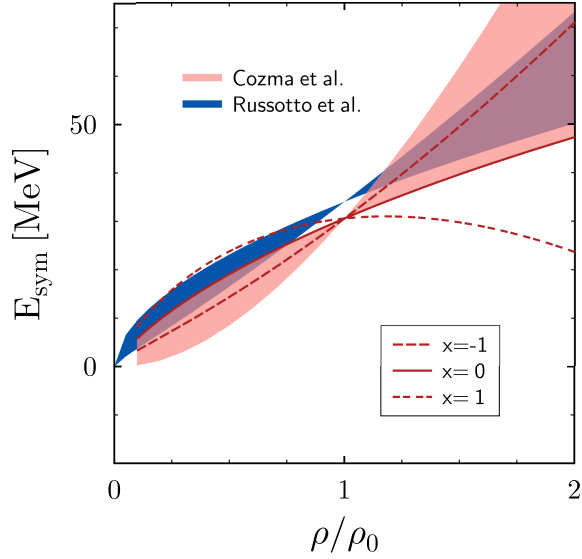}}
   \caption{(Color online) Constraints on the density dependence of the symmetry energy obtained by Cozma {\it et al.} from 
comparing predictions of the T\"{u}QMD for the neutron-proton elliptic-flow difference and ratio
to FOPI-LAND experimental data (Ref.~\protect\cite{cozma13}). 
The result of Russotto {\it et al.} (Ref.~\protect\cite{russotto11}) is also shown, 
together with the Gogny-inspired MDI parametrization of the symmetry energy for three values of 
the stiffness parameter: x = -1 (stiff), x = 0, and x = 1 (soft)
(from Ref.~\protect\cite{cozma13}; Copyright (2013) by the American Physical Society).
}
\label{fig:cozma}
\end{figure}

A more comprehensive study of parameter dependencies in connection with the FOPI-LAND data was performed by 
Cozma {\it et al.} with the T\"{u}QMD model~\cite{cozma13}. The effects of the selected microscopic nucleon-nucleon 
cross-sections, of the compressibility of symmetric nuclear matter, of the optical potential, and of the parametrization 
of the symmetry-energy were thoroughly studied. 
The most significant contributions to the uncertainty of the slope parameter $L$ were found to be caused by 
the compressibility $K_0$ of symmetric matter and the width $L$ of the nucleon wave function used 
in the calculations. Only small changes occurred when switching from the power-law parametrization 
to the Gogny inspired MDI force with momentum dependence in the asymmetric part of the interaction.

\begin{figure}[ht]           
  \centerline{\includegraphics[width=250pt]{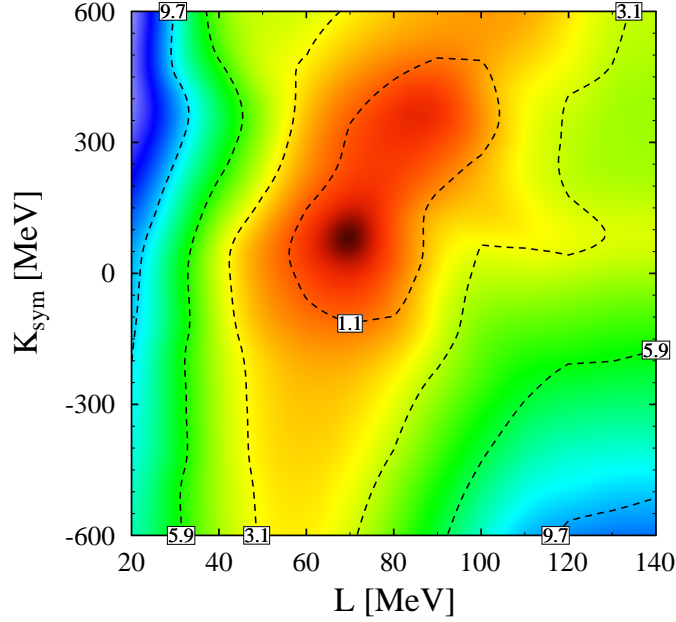}}
   \caption{(Color online) Constraints for the slope $L$ and the curvature $K_{\rm sym}$ extracted from
the FOPI-LAND and ASY-EOS data sets. Curves for the 1, 2, 3, and 4 $\sigma$ confidence levels are shown
labeled by the corresponding $\chi^2$/point
(from Ref.~\protect\cite{cozma18}, reprinted with kind permission from Springer Science+Business Media).
}
\label{fig:cozma2}
\end{figure}

In  Fig.~\ref{fig:cozma}, the explicit constraints on the density dependence
of the symmetry energy obtained in this study from the comparison of theoretical and 
experimental values of elliptic-flow ratios and differences
are presented. The result of Ref.~\cite{russotto11} obtained with the UrQMD parametrization (Eq.~\ref{eq:pot_term}) is also 
given in the figure. 
The two studies employ independent flavors of the QMD transport model (T\"{u}QMD vs. UrQMD) and parametrizations of both, 
the isoscalar and isovector EoS that differ. The constraints on the density dependence of 
the symmetry energy obtained with these different ingredients are shown to be in agreement with
each other. The error band of the result of Cozma is wider because the consequences of the wider parameter space that 
has been tested are included.
Both estimates support a moderately stiff to linear density dependence 
corresponding to a parametrization $x=-1.0 \pm 1.0$. This result is shown to be incompatible with
the supersoft density dependence of the symmetry energy that was obtained in a
first analysis~\cite{xiao09} of the FOPI $\pi^-/\pi^+$  ratios~\cite{reisdorf07}. More recent studies indicate, however,
that the interpretation of the pion ratios cannot be considered as 
conclusive at the present time (see, e.g., Refs.~\cite{guo2014,cozma16,lili2017,trautwolt17,zhangko2018} and references 
given therein). 

\begin{figure}[ht]           
  \centerline{\includegraphics[width=260pt]{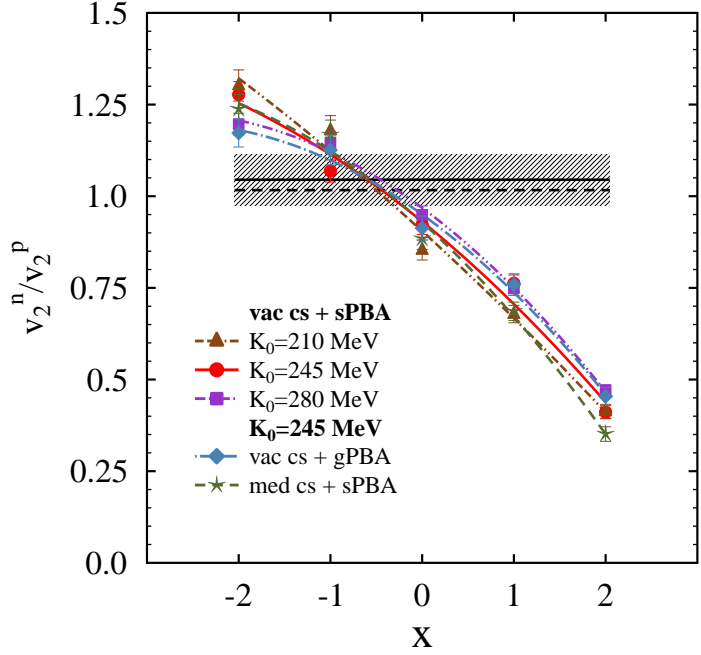}}
   \caption{(Color online) Neutron-to-proton elliptic flow ratio as a function of the stiffness parameter $x$ in
comparison with the FOPI-LAND data. Experimental results as reported in Refs.~\protect\cite{wang14} and 
\protect\cite{cozma13} are indicated by the full and dashed horizontal lines, respectively.
Only the error band corresponding to the former data set (hashed) is shown in the figure
(from Ref.~\protect\cite{cozma18}, reprinted with kind permission from Springer Science+Business Media).
}
\label{fig:cozma3}
\end{figure}

A major step forward was made by Cozma very recently~\cite{cozma18}. A density dependent term was added to the MDI force,
leading to a new force MDI2 with an additional parameter $y$ that, together with the original $x$ of MDI,
permitted choosing the slope and curvature parameters $L$ and $K_{\rm sym}$ of the symmetry energy independently. By 
lifting the correlation of these two parameters, present in most models, uncertainties in the short-range three-body 
force are admitted and thereby taken into account~\cite{cozma18}. The crossing 
of the various options was moved from the saturation density to the density 0.1~fm$^{-3}$ and placed at 25.5 MeV, i.e.
at the value obtained by Brown~\cite{brown13}. The T\"{u}QMD supplemented by a phase-space coalescence model fitted to 
FOPI experimental multiplicities of free nucleons and light clusters~\cite{reisdorf12} was used to compare the 
theoretical predictions with the FOPI-LAND neutron-to-proton and ASY-EOS neutron-to-charged-particles elliptic flow ratios.
The correlation map obtained for $L$ and $K_{\rm sym}$ is shown in 
Fig.~\ref{fig:cozma2}. The correlation peaks at $L = 72$~MeV and $K_{\rm sym} = 78$~MeV. Additional corrections 
that were applied are motivated and described in the paper. The final result presented there is 
$L = 85 \pm 22({\rm exp}) \pm 20 ({\rm th}) \pm 12 ({\rm sys})$~MeV and 
$K_{\rm sym} = 96 \pm 315({\rm exp}) \pm 170 ({\rm th}) \pm 166 ({\rm sys})$~MeV~\cite{cozma18}. The systematic errors
listed separately take, e.g., into account that the experimental light-cluster-to-proton multiplicity ratios are 
underestimated by the QMD model. 

The largest contribution of the error is experimental and, as it turns out, mainly related to the neutron-proton flow 
ratio presently only available from the FOPI-LAND data and thus affected by the limited statistics collected in that experiment. 
This can be improved with new measurements, so that a much more precise value for the curvature term can 
be expected. The example shown in Fig.~\ref{fig:cozma3} illustrates that point, in that case for calculations with 
the original MDI force. It shows that various choices made for the symmetric-matter compressibility $K_0$, the cross 
section parametrizations, and details of the algorithm controlling the observation of Pauli blocking have only small 
consequences for the calculated ratios. Also a more precise constraint for $x$ can thus be expected from new measurements.

\section{NEUTRON STAR RADII}

The results of the analysis of the GW170817 event has mostly been presented in the form of predictions for the radius 
of a standard neutron star with a mass equal to 1.4 solar masses. Bauswein {\it et al.} were among the first to realize the 
consequences of the observation of the electromagnetic emissions following the merger. It allowed them to conclude that a prompt 
collapse of the remnant cannot have occurred~\cite{bausw2017}. 
Based on that fact, a lower limit of 10.68~km was deduced for the radius of a nonrotating 
neutron star of 1.6 solar masses, with the error interval extending down to 10.64~km. Fattoyev {\it et al.} compared the 
tidal deformability parameter $\tilde{\Lambda} = 800$ reported by LIGO and Virgo in 2017~\cite{abbott2017} with predictions 
of a selected set of equations of 
state~\cite{fatto2018}. They concluded that the observed deformability defines a limit on the stiffness of the equation 
of state corresponding to $R_{1.4} \le 13.76$~km for a neutron star with 1.4 solar masses. This interval was narrowed down
to $R = 11.9 \pm 1.4$~km in the revised analysis of the LIGO and Virgo collaborations \cite{abbott2018}. 
It reduced the upper upper limit to $R \le 13.3$~km for a star with approximately 1.4 solar masses.

In the following, the tight correlation of the pressure $p_0$ with $R_{1.4}$ presented by Lattimer and 
Prakash \cite{lattprak2016} is used to determine the radii corresponding to the pressures determined in the 
elliptic-flow experiments. From the slope parameter $L=72 \pm 13$~MeV obtained with the UrQMD parametrization 
(Fig.~\ref{fig:russotto_fig18}), the pressure $p_0 = 3.8 \pm 0.7$~MeV was deduced~\cite{russotto16}. It corresponds 
to a radius interval $11.9 \le R_{1.4} \le 13.4$~km. With the lower limit $L= 52$~MeV, obtained by using 
$E_{\rm sym}(\rho_0) = 31$~MeV for the parametrization of the symmetry energy, the lower limit is reduced to 11.6~km. 
It is still about 1~km larger 
than the lower ends of the intervals arising from the GW170817 analyses. The upper limits of around 13.5~km are all 
consistent with each other.

The analysis of Cozma based on the MDI2 force and the ASY-EOS and FOPI-LAND data yields a slope parameter 
with a considerably larger error interval as a result of considering a nearly complete list of model 
dependencies and of taking their associated uncertainties into account~\cite{cozma18}. A unique feature is the independent 
variation of the slope and curvature parameters, $L$ and $K_{\rm sym}$, parameters that appear correlated in practically all 
descriptions of the nuclear mean field in transport models. 
Combining the experimental, theoretical and systematic errors quadratically leads to $L=85 \pm 32$~MeV, 
i.e. to a lower limit $L=53$~MeV. It is practically identical to the UrQMD result and leads to the same lower limit 
$R_{1.4} \ge 11.6$~km. The upper part of the interval extends beyond the GW170817 limits and, with $L>85$~MeV, 
enters a region not compatible with the unitary-gas limit discussed at this conference~\cite{lattimer2019,tews2016}.

A tight upper limit $L \le 70$~MeV has recently been presented by Zenihiro {\it et al.}, deduced from a 
measurement of the neutron skin of $^{48}$Ca with elastic scattering of polarized protons at an incident 
energy 295~MeV~\cite{zenihiro2018,sagawa2019}. 
The deduced interval $20 \le L \le 70$~MeV includes small values of the slope parameter but overlaps with the elliptic-flow 
results, emphasizing the lower part of the error band with its limit $L \ge 53$ obtained by Cozma~\cite{cozma18}. 
The authors argue that a determination relying on a simple reaction mechanism and independent of a specific nuclear structure 
model is very important to examine the theory. A similar strategy is followed by Aumann {\it et al.} with their recent proposal 
for experiments at FAIR~\cite{aumann2017}.
The correlation of the obtained neutron skin thickness $\Delta R_{np} =0.17$~fm of $^{48}$Ca with $L$ adopted by 
Zenihiro {\it et al.} agrees precisely with that reported by Brown in his recent analysis of charge radii of 
mirror nuclei~\cite{brown2017}. The upper limit $L \le 70$~MeV corresponds to a radius $R_{1.4} \le 12.6$~km. 

A compilation of radius estimates based on analyses of the GW170817 data is available in Table II of the preprint of 
Montana {\it et al.}~\cite{montana2018}. Nearly all of the presented ranges encompass the
11.6 to 12.6~km interval that follows from the two laboratory experiments measuring the elliptic flows in $^{197}$Au+$^{197}$Au
and the neutron skin of $^{48}$Ca. It holds also for the analysis of Tews {\it et al.}~\cite{tews2018} presented 
at this conference~\cite{margueron2019}. A tighter lower limit for the deformability $\tilde{\Lambda}$ has been 
reported by Radice {\it et al.}~\cite{radice2018}. On the basis of their 
simulations and the current interpretation of the UV/optical/infrared data collected in connection with GW170817, they 
conclude that values of $\tilde{\Lambda}$ smaller than 400 are tentatively excluded.
It corresponds to radii of 12 km or larger according to the systematics established by 
Burgio {\it et al.}~\cite{burgio18,wei2019}. The very similar result obtained by Malik {\it et al.}~\cite{malik2018} 
has been reported at this conference~\cite{providencia}.  
 
The quality of the agreement of neutron star radii inferred from 
terrestrial and celestial data can be considered as very promising. Further studies and new measurements of high precision 
will permit critical assessments of the applied 
methods and models. The neutron star radius appears as a very useful basis for comparisons even though none of the 
above-mentioned studies 
have directly measured a radius. Radii determined from X-ray observations as compiled in the review of \"{O}zel and Freire 
fall into the interval 9.9 - 11.2 km for 1.4 solar mass stars which is marginally consistent with the above 
estimate~\cite{oezel2016}. The earlier result 11 - 12 km of Steiner {\it et al.} based on X-ray data from six neutron 
stars is closer and partly overlaps with it~\cite{steiner2010,steiner2012}.
It will thus be extremely interesting to see first results from the NICER campaign on the International Space 
Station~\cite{NICER}. Several neutron stars have apparently been observed and their compactness is presently being determined 
from the measured pulse profiles~\cite{watts2019}. 

\section{FUTURE OPPORTUNITIES}

The experimental results and the theoretical study of the density range that has been probed provide a strong 
encouragement for continuing the flow measurements with improved detection systems. It is evident from Fig.~\ref{fig:cozma3} 
that a more precise 
measurement of the neutron-proton elliptic flow ratio, even at the same energy, will strongly reduce the error margins 
of the analyses presented 
here. It requires isotopic resolution for light charged particles that was achieved in the FOPI-LAND experiment, 
however at the cost of significant cuts into the counting statistics~\cite{lamb94}. Future experiments will, 
therefore, benefit from the improved capabilities of the NeuLAND detector
presently constructed as part of the $R^{3}B$ setup at FAIR~\cite{NeuLAND}. Model studies, furthermore, 
indicate that the sensitivity to the stiffness of the symmetry energy is still significant at incident 
energies as high as 800~MeV/nucleon or 1 GeV/nucleon~\cite{russotto2017,lukasik2018}. 
The sensitivity to density increases slowly but can be expected to be centered near 
twice saturation at these energies~\cite{lefevre16}. Energies lower than 400~MeV/nucleon have also been suggested as 
useful for the determination of the curvature term in the asymmetric-matter EoS~\cite{cozma18}. 
With measurements at several energies, as proposed to be conducted at FAIR~\cite{russotto2017}, the analysis within 
transport-model descriptions will be more stringently tested.

\begin{figure}[ht]           
  \centerline{\includegraphics[width=235pt]{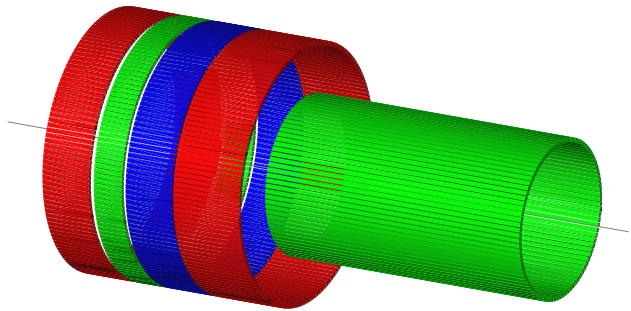}}
   \caption{(Color online) Schematic design of the KRAk\'{o}w Barrel (KRAB). Altogether 736 elements of fast scintillation 
fiber detectors are arranged in five cylindrical barrels concentric with the beam axis and covering polar angles 
$\theta_{\rm lab}$ from 30$^{\circ}$ to 165$^{\circ}$. The beam enters from the right through the narrow cylinder located 
upstream of the target plane. The four rings with larger diameter follow downstream of the target 
plane (from Ref.~\protect\cite{lukasik2018}). 
}
\label{fig:krab}
\end{figure}

A dedicated detection system named KRAk\'{o}w Barrel (KRAB) is presently under construction~\cite{lukasik2018}.
Its design is based on the experience gained during the ASY-EOS campaign and the subsequent data analysis~\cite{russotto16}. 
It consists of an arrangement of 736 fast scintillation fibers surrounding the target and is meant to replace the Microball 
used in the ASY-EOS experiment (cf. Fig.~\ref{fig:setup}). With its high granularity, it will permit superior measurements 
of the azimuthal particle distributions and of the charged-particle multiplicities. 
The observed quality of the simulated signals gives rise to the expectation that the device will indeed provide
very precise information on the orientation of the reaction plane and on the centrality of
the collision. It will also play an important role by suppressing background reactions occurring in
materials other than the target.
The KRAB detector will cover about 85\% of the total solid angle. Together with a suitable
forward-wall type detection system, more than 90\% of the $4\pi$ solid angle can be subtended without interfering with the 
detection of neutrons in lateral directions.

\section{CONCLUSION AND OUTLOOK}

Several analyses of the FOPI-LAND and the more recent ASY-EOS elliptic-flow ratios agree on the moderately soft to linear
density dependence of the symmetry energy. The original result $L = 83 \pm 26$~MeV obtained with the UrQMD transport model 
and the FOPI-LAND data~\cite{russotto11} has been confirmed by Wang {\it et al.}~\cite{wang14}
using selected Skyrme forces within the UrQMD ($L = 89 \pm 45$~MeV, 2-$\sigma$ uncertainty) and, with higher precision, 
by the ASY-EOS result $L = 72 \pm 13$~MeV obtained with the UrQMD transport model~\cite{russotto16}. Cozma's result, 
$L = 85 \pm 22({\rm exp}) \pm 20 ({\rm th}) \pm 12 ({\rm sys})$~MeV, has a larger error because the
full parameter space, including the isoscalar sector, has been explored in the calculations~\cite{cozma18}. It is unique 
because the correlation of the slope and curvature parameters that is present in most models has been lifted by using 
the MDI2 force in this particular version of the T\"{u}QMD
transport model. The obtained slope parameters are all compatible 
with each other. They are slightly larger than the weighted averages $L \approx 59$~MeV calculated by Li and 
Han~\cite{lihan2013} and Oertel {\it et al.}~\cite{oertel2017} by averaging over many terrestrial and astrophysical 
studies but touch this value with the lower end of their 1-$\sigma$ 
error margins. The significance of the lower bound of the error interval in comparison with
the limits obtained from the GW170817 observations is emphasized.

The sensitivity study shows that suprasaturation densities are effectively probed with the elliptic flow ratio or 
difference of neutrons with respect to charged particles. Because the interpretation of the FOPI pion ratios~\cite{reisdorf07} 
is not yet conclusive (see, e.g., Refs.~\cite{guo2014,cozma16,lili2017,trautwolt17,zhangko2018} and 
references given therein), the differential flow observables may be considered as presently unique as a terrestrial 
source of information for the EoS of asymmetric matter at high densities. The different sensitivities probed by flow
ratios of neutrons with respect to protons and with respect to light charged particles provide access to the
curvature parameter $K_{\rm sym}$. Determined with the existing data, the uncertainty of $K_{\rm sym}$ is still large but 
errors close to $\pm 150$~MeV may be ultimately  within reach~\cite{cozma18}.
The presented experimental results and theoretical studies thus provide a strong 
encouragement for the continuation of flow measurements of the present kind with improved detection systems at FAIR,
the Facility for Antiproton and Ion Research~\cite{russotto2017,lukasik2018}.

The discussion has shown that useful comparisons of results derived from laboratory experiments and from 
astrophysical observations can be made in the density interval of once to twice the saturation value. 
The pressures deduced from the elliptic flow measurements of the ASY-EOS and FOPI Collaborations are fully compatible 
with the pressures in that interval derived with a parametric set of equations of state by the LIGO and Virgo 
Collaboration~\cite{abbott2018}. The tight correlation of the pressure of neutron matter at saturation with the radius of 
a 1.4 solar-mass neutron star, as reported by Lattimer and Prakash~\cite{lattprak2016} and similarly by Newton 
and Li~\cite{newtonli2009}, permits predictions for the radii. 

With the upper limit taken from the recent 
measurement of the $^{48}$Ca neutron skin, the interval of $11.6 \le R_{1.4} \le 12.6$~km was obtained as the present 
prediction based on laboratory experiments. It is comfortably placed within the interval $11.9 \pm 1.4$~km of the refined 
analysis of the LIGO and Virgo observations and fully compatible with the radius 11.9 km that corresponds to the  
average values $L = 58.9$~MeV and $L = 58.7$~MeV reported in Refs.~\cite{lihan2013,oertel2017}. It overlaps with the narrow 
interval of about 12 km to 13.3 km 
corresponding to the results obtained by Radice {\it et al.} and Malik {\it et al.} from simulations based on the GW170817 
data ~\cite{radice2018,malik2018,providencia}. Radii deduced from X-ray observations are slightly smaller; 9.9 to 11.2 km is 
the interval reported by \"{O}zel and Freire~\cite{oezel2016} but the earlier result 11 to 12 km of Steiner {\it et al.} 
partly overlaps with the terrestrial predictions~\cite{steiner2010,steiner2012}. 
Obviously, the radii expected to be soon released by the NICER Collaboration will be extremely 
interesting~\cite{NICER,watts2019}.\\

{\bf Acknowledgment} 

The author is indebted to M.~D.~Cozma, A.~Le~F\`{e}vre, Y.~Leifels, J.~{\L}ukasik, P.~Russotto, Yongjia Wang, and H.~H.~Wolter 
for fruitful discussions in connection with the preparation of this talk. The very cooperative spirit 
within the ASY-EOS Collaboration is gratefully acknowledged (see Ref.~\cite{russotto16} for the complete list of authors).

\end{document}